\documentclass[aps,prb,twocolumn,superscriptaddress,floatfix]{revtex4-1}

\usepackage{graphicx,graphics}
\usepackage{dcolumn}
\usepackage{amsmath,amssymb,amsfonts}
\usepackage{latexsym,verbatim}
\usepackage{bm}
\usepackage{color}
\usepackage{ulem}
\usepackage[percent]{overpic}
\usepackage[breaklinks=true,colorlinks,citecolor=blue,linkcolor=blue,urlcolor=blue]{hyperref}

\DeclareMathAlphabet\mathbfcal{OMS}{cmsy}{b}{n}

\newcommand{\bp}{{\bm p}}
\newcommand{\bq}{{\bm q}}
\newcommand{\br}{{\bm r}}

\newcommand{\epsF}{\varepsilon_{\rm F}}
\newcommand{\vD}{v_{\mathrm{D}}}

\begin{document}

\title{Helicons in Weyl semimetals}
\author{Francesco M.D. Pellegrino}
\email{francesco.pellegrino@sns.it}
\affiliation{NEST, Scuola Normale Superiore, I-56126 Pisa,~Italy}
\author{Mikhail I. Katsnelson}
\affiliation{Radboud University, Institute for Molecules and Materials, NL-6525 AJ Nijmegen,~The
Netherlands}
\author{Marco Polini}
\affiliation{Istituto Italiano di Tecnologia, Graphene Labs, Via Morego 30, I-16163 Genova,~Italy}

\begin{abstract}
Helicons are transverse electromagnetic waves propagating in three-dimensional (3D) 
electron systems subject to a static magnetic field. In this work we present a theory of helicons propagating through a 3D Weyl semimetal. Our approach relies on the evaluation of the optical conductivity tensor from semiclassical Boltzmann transport theory, with the inclusion of certain Berry curvature corrections that have been neglected in the earlier literature (such as the one due to the orbital magnetic moment). 
We demonstrate that the axion term characterizing the electromagnetic response of Weyl semimetals dramatically alters the helicon dispersion with respect to that in non-topological metals. We also discuss axion-related anomalies that appear in the plasmon dispersion relation.
\end{abstract}

\maketitle

\noindent {\it Introduction.---}After lightning strikes, very low-frequency transverse electromagnetic waves called ``whistlers''~\cite{jackson} propagate in the ionosphere, from one hemisphere to the other, along the Earth's magnetic field lines. Interestingly, whistlers have a solid-state analog, which is usually called ``helicons"~\cite{konstantinov_jetp_1960,aigrain,bowers_prl_1961,maxfield_amjp_1969,Platzman_and_Wolff}. These transverse electromagnetic waves propagate in three-dimensional (3D) uncompensated metals subject to a uniform static magnetic field ${\bm B} = B \hat{\bm z}$. Helicons ultimately stem from the existence of the cyclotron resonance~\cite{Platzman_and_Wolff}, a single-particle excitation of the 3D electron system occuring at the cyclotron frequency $\omega_{\rm c} = eB/(m c)$, where $m$ ($-e$) is the electron mass (charge) and $c$ is the speed of light in vacuum. In ordinary metals, helicons propagating along the magnetic field direction with a wave vector ${\bm q} = q \hat{\bm z}$  
have the following free-particle-like dispersion relation~\cite{Platzman_and_Wolff},
\begin{equation}\label{eq:standard_helicon}
\Omega_{\rm h}(q) = \frac{\hbar q^2}{2 m_{\rm h}} \ll \omega_{\rm c}~,
\end{equation}
where $m_{\rm h} = \hbar \omega_{\rm p}^2/(2 \omega_{\rm c}c^2)\propto 1/B$ is the helicon effective mass and $\omega^2_{\rm p} = 4\pi n_{\rm e} e^2/m$ the usual 3D plasmon frequency~\cite{Pines_and_Nozieres,Giuliani_and_Vignale}. Helicons are relatively useless as a probe of many-body effects in the metallic medium but are interesting since they can hybridize with sound waves or display interesting damping behaviors when the magnetic field direction is tilted from their propagation direction~\cite{Platzman_and_Wolff}.

In this work we demonstrate that the helicon dispersion in Weyl semimetals (WSMs)~\cite{wan_prb_2011,burkov_prl_2011,singh_prb_2012,weng_prx_2015,huang_naturecommun_2015} is greatly altered with respect to the textbook result (\ref{eq:standard_helicon}). In particular, we show that helicon propagation in WSMs carries precious information on the space- (${\bm b}$) and time-like ($b_0$) components of the axion angle. We also highlight axion anomalies in the plasmon sector, which, to the best of our knowledge, have not yet been noticed.

\noindent {\it Maxwell equations in WSMs.---}WSMs are recently discovered~\cite{huang_arxiv_2015,lv_prx_2015,lv_arxiv_2015a,xu_prx_2015,xu_science_2015,ghimire_jpcd_2015} 3D topological metals displaying an intriguing electromagnetic response and Fermi-arc surface states. For the sake of simplicity, we here consider the Hamiltonian of a WSM with {\it two} nodes only~\cite{burkov_prl_2011}:
\begin{equation}\label{eq:WeylHamiltonian}
{\cal H} =   \hbar  v_{\rm D}\tau^z {\bm \sigma} \cdot\left(-i \nabla +\tau^z  {\bm b}\right) +\hbar \tau^z b_0~.
\end{equation}
Here, $v_{\rm D}$ is the Dirac-Weyl velocity, $\tau^z$ describes the node degree of freedom with chirality $\pm 1$, and the 3D vector of Pauli matrices $\bm \sigma=(\sigma^x,\sigma^y,\sigma^z)^{\rm T}$ describes conduction- and valence-band degrees of freedom. The two Weyl nodes are located at $\pm {\bm b}$ and shifted by $2 \hbar b_0$ in energy. Our results below can be easily generalized to the case of more than two Weyl nodes. Furthermore, in the present work we study helicons (and plasmons) in bulk Weyl semimetals, while the interplay between finite-thickness effects in a slab geometry (such as Fermi arcs) and electromagnetic wave propagation will be the scope of future works.

It has been demonstrated~\cite{zyuzin_prb_2012} that the terms proportional  to 
$b_0$ and ${\bm b} = (b_x,b_y,b_z)^{\rm T}$ in Eq.~(\ref{eq:WeylHamiltonian}) can be gauged away. After this transformation, the Hamiltonian reduces to ${\cal H} = -i \hbar  v_{\rm D}\tau^z {\bm \sigma} \cdot \nabla$. 
Because of the chiral anomaly~\cite{peskin}, however, the aforementioned gauge transformation generates an additional term in the Lagrangian ${\cal L}_{\rm em}$ that describes the coupling between light and 3D WSMs~\cite{zyuzin_prb_2012}:
\begin{equation}\label{eq:emLagrangian}
{\cal L}_{\rm em}=\frac{1}{8 \pi} ({\bm E}^2-{\bm B}^2)-\rho \phi + {\bm J} \cdot {\bm A} +  {\cal L}_{\theta}~,
\end{equation}
where
\begin{equation}\label{eq:axionterm}
 {\cal L}_{\theta} = -\frac{\alpha}{4 \pi^2}  \theta (\br, t) {\bm E}\cdot {\bm B}~.
\end{equation}
Here, $\alpha=e^2/(\hbar c) \simeq 1/137$ is the usual QED fine-structure constant 
and $\theta (\br, t)\equiv 2 ({\bm b} \cdot \br -  b_0  t)$ is the so-called axion angle.
The additional axion term ${\cal L}_{ \theta}$ 
changes two of the four Maxwell equations, i.e.~\cite{wilczek_prl_1987}
\begin{equation}\label{eq:nablaE}
 \nabla \cdot {\bm E} = 4 \pi \left(\rho + \frac{\alpha}{2 \pi^2} {\bm b} \cdot {\bm B}\right)
\end{equation}
and
\begin{equation}\label{eq:nablaB}
-\frac{1}{c} \frac{\partial {\bm E}}{\partial t} + \nabla \times {\bm B} =  \frac{4 \pi}{c} \left(  {\bm J} 
-\frac{\alpha}{2 \pi^2}  {c \bm b} \times {\bm E} + \frac{\alpha}{2 \pi^2}   b_0 {\bm B}
\right)~.
\end{equation}
Faraday's law, $\nabla \times {\bm E} = - c^{-1} \partial {\bm B}/\partial t$, and the equation stating the absence of free magnetic poles, $\nabla \cdot {\bm B} = 0$, are unchanged.

Eliminating ${\bm B}$, we obtain the following modified wave equation in a WSM:
\begin{eqnarray}\label{eq:wave}
-\frac{1}{c^2} \frac{\partial^2 {\bm E}}{\partial t^2} &-& \nabla \times (\nabla \times {\bm E}) =
\frac{4 \pi}{c^2}  \frac{\partial {\bm J}}{\partial t} \nonumber\\
&-&\frac{2 \alpha}{\pi c} {\bm b}\times  \frac{\partial {\bm E}}{\partial t} 
-\frac{2\alpha }{\pi c} b_0 \nabla \times {\bm E}~.
\end{eqnarray}
As usual, we now need an expression that relates the current ${\bm J}$ to the electric field ${\bm E}$, which we  proceed to derive by utilizing a semiclassical approach.

\noindent {\it Semiclassical Boltzmann transport theory in WSMs.---}In the linear response regime~\cite{Platzman_and_Wolff,Pines_and_Nozieres,Giuliani_and_Vignale}, the Fourier components ${\bm J}({\bm q}, \omega)$ of the induced current density are linearly dependent on the total electric field (i.e.~the sum of the external field and the Hartree contribution), i.e.~$J_m(\bq, \omega) =\sum_n \tilde{ \sigma}_{m n}(\bq,\omega)E_n (\bq, \omega)$, 
where $\tilde{\sigma}_{m n}(\bq,\omega) $ is the optical conductivity tensor and the indices $m, n$ run over the Cartesian coordinates $x$, $y$, and $z$.
We will work in the {\it local} approximation~\cite{Platzman_and_Wolff}, which is justified in the limit $q \ll R^{-1}_{\rm c}$, where $R_{\rm c} = v_{\rm D}/\omega_{\rm c}$ is the cyclotron radius, with 
$\omega_{\rm c}= e B/(m_{\rm c} c)$ the cyclotron frequency, $m_{\rm c}=\epsF/\vD^2$ the WSM cyclotron mass, and $\epsF$ the Fermi energy. We therefore have $\tilde{ \sigma}_{m n}(\bq,\omega) \approx \tilde{\sigma}_{m n}({\bm 0}, \omega)\equiv \delta_{m n} \sigma_{\rm b}(\omega)+\sigma_{m n}(\omega)$, where $\sigma_{\rm b}(\omega)=-i  (\epsilon_{\rm b}-1) \omega/(4 \pi)$ is the bound-charge contribution, while $\sigma_{m n}(\omega)$ represents the free-charge contribution~\cite{Grosso_and_PastoriParravicini}.
The latter quantity can be calculated by utilizing a semiclassical Boltzmann transport approach, which is justified when $\omega, \omega_{\rm c} \ll \epsF/\hbar$. Also, we focus on the collisionless $\omega \tau \gg 1$ regime, where $\tau = {\rm min}(\tau_{\rm intra}, \tau_{\rm inter})$ is the {\it shortest} between intra-node $\tau_{\rm intra}$ (e.g.~due momentum-non-conserving collisions) and inter-node $\tau_{\rm inter}$ scattering times.

Our interest in this work in on doped Weyl semimetals ($k_{\rm B} T\ll \epsF$, where $T$ is temperature). The situation for $k_{\rm B} T\gg \epsF$ (neutral Weyl semimetals) is much more complicated as one needs to include finite temperature effects (thermally excited carriers) and disorder.

For a given chirality $g=\pm$ of a single Weyl node, the semiclassical Boltzmann equation (SBE) reads as following~\cite{son_prb_2013,kim_prb_2014}:
\begin{equation}\label{eq:semiclassicalBE}
 \frac{\partial f_{g}}{\partial t}+ \dot{\bp}_g\cdot \nabla_{\bp} f_{g}+ \dot{\br}_g\cdot \nabla_{\br} f_{g}=0~.
\end{equation}
Here, $f_{g}$ is the electron distribution function. 
In the presence of a static magnetic field ${\bm B}$ and a time varying electric field ${\bm E}$, 
the semiclassical equations of motion are~\cite{dixiao_rmp_2010}
\begin{equation}\label{eq:dot_r}
 \dot{\br}= {\bm v}_g(\bp)-\dot{\bp} \times {\bm \Omega}_g(\bp)
\end{equation}
and
\begin{equation}\label{eq:dot_p}
\dot{\bp} = -\frac{e}{\hbar} {\bm E}-\frac{e}{\hbar c} \dot{\br}\times {\bm B}~. 
\end{equation}
The first term on the right-hand-side of Eq.~(\ref{eq:dot_r}) is ${\bm v}_g(\bp)=\hbar^{-1}\nabla_{\bp} \varepsilon_{g}(\bp)$, defined in terms of an effective band dispersion $\varepsilon_{g}(\bp)$. In topological metals such as WSMs, this quantity acquires a term due to the intrinsic orbital moment~\cite{dixiao_rmp_2010}, i.e.~$\varepsilon_g(\bp)=\varepsilon_0(\bp)-{\bm m}_g(\bp)\cdot {\bm B}$, where
 $\varepsilon_0(\bp)=\hbar \vD p$ with $p = |\bp|$ is the ordinary conduction band energy while ${\bm m}_g(\bp)$ is the orbital moment~\cite{dixiao_rmp_2010}, i.e.~${\bm m}_g(\bp) =  \gamma e \varepsilon_0(\bp) {\bm \Omega}_g(\bp)/(\hbar c)$. Here, $\gamma$ is a dimensionless control parameter and ${\bm \Omega}_g(\bp)=-g  \bp/(2 p^3)$ is the WSM Berry curvature~\cite{dixiao_rmp_2010}. The parameter $\gamma$ takes two values: $\gamma =1$ is what one should use, while $\gamma = 0$ is what one should use to {\it artificially} discard the impact of the orbital magnetic moment.

Using Eqs.~(\ref{eq:dot_r})-(\ref{eq:dot_p}) and carrying out straightforward algebraic manipulations we find:
\begin{equation}
\dot{\br} = \dot{\br}_{B}+  \dot{\br}_{E}~,
\end{equation}
 where
\begin{equation}
\dot{ \br}_{B}\equiv D^{-1}_g(\bp)\left[ {\bm v}_g(\bp)+\frac{e }{ \hbar c} {\bm \Omega}_g(\bp)\cdot {\bm v}_g(\bp)  {\bm B}\right]~,
\end{equation}
and
\begin{equation}
\dot{ \br}_{E}\equiv D^{-1}_g(\bp) \frac{e}{\hbar} {\bm E} \times {\bm \Omega}_g(\bp)~.
\end{equation}
Here, $D_g(\bp)\equiv [1+ e {\bm \Omega}_g\cdot {\bm B}/(\hbar c)]$ and the group velocity ${\bm v}_g(\bp)$ is given by
\begin{equation}
{\bm v}_g(\bp)=\vD \hat{\bp}\left[ 1+\gamma \frac{2e}{\hbar c} {\bm \Omega}_g(\bp)\cdot {\bm B} \right]
-\gamma\frac{e \vD}{\hbar c}  \Omega_g(\bp) {\bm B}~,
\end{equation}
with ${\hat \bp}=\bp/p$ and $\Omega_g(\bp)=|{\bm \Omega}_g(\bp)|$. Similarly, we find: $\dot{\bp} = \dot{\bp}_{B}+  \dot{\bp}_{E}$ with 
$\dot{\bp}_{B}\equiv -D^{-1}_g(\bp) e {\bm v}_g(\bp) \times {\bm B}/c$ and 
$\dot{\bp}_{E} \equiv -D^{-1}_g(\bp)\left[ e {\bm E} + e({\bm E} \cdot {\bm B}){\bm \Omega}_g(\bp)/(\hbar c)\right]$.

Let us start by setting ${\bm E} ={\bm 0}$ while keeping ${\bm B}=B \hat{\bm z}$ finite. 
In this case, the SBE (\ref{eq:semiclassicalBE}) is solved by
\begin{equation}\label{eq:FD}
f_g^{(0)}(\bp) \equiv\frac{1}{\exp \left[\displaystyle \frac{\varepsilon_g(\bp)-\epsF}{k_{\rm B}T}\right]+1}~,
\end{equation}
when in the collision integral ${\cal I}_g$ we take $f_{\rm eq}=f^{(0)}_g(\bp)$. We now want to solve the SBE up to first order in the amplitude of a homogeneous time-dependent electric field, ${\bm E} = \tilde{\bm E}( \omega) e^{-i   \omega t }$. To this end, it is useful~\cite{orlita_prl_2012} to exploit the symmetry of system by using cylindrical coordinates: $\bp\equiv(\sqrt{p^2-p_z^2}\cos(\varphi), \sqrt{p^2-p_z^2}\sin(\varphi),p_z)^{\rm T}$. We seek a solution of the SBE of the form
\begin{equation}\label{eq:linearization}
f_g(\bp,t)=f_g^{(0)}(\bp)+\delta f_{g}(\bp,t)~,
\end{equation}
where $\delta f_{g}(\bp,t)$ is linear in $\tilde{\bm E}$ and is parametrized as following: 
\begin{equation}\label{eq:deltaf}
\delta f_{g}(\bp,t)=-\frac{\partial  f^{(0)}_{g}}{\partial \varepsilon_{g}}\left(X_- e^{i \varphi} +X_+ e^{-i \varphi} +
 X_0\right) e^{-i \omega t}~,
\end{equation}
with $X_{\pm,0} = X_{\pm, 0}(p,p_z)$. The linearization of the SBE (\ref{eq:semiclassicalBE}) is greatly simplified by the observation that $p_{z}$ and $p$ are constants of the motion 
in the limit $\tilde{\bm E} \to {\bm 0}$.

Inserting (\ref{eq:linearization})-(\ref{eq:deltaf}) in Eq.~(\ref{eq:semiclassicalBE}) we find
\begin{equation}\label{eq:xpluminus}
 X_\pm=e \vD \delta \frac{\displaystyle 1-\gamma g\frac{e}{\hbar c} \frac{p_z}{p^3} B  }{\displaystyle1-g\frac{e}{2\hbar c} \frac{p_z}{p^3} B  }
\frac{ \sqrt{p^2-p_z^2}}{2p} \frac{\tilde{E}_x \pm i \tilde{E}_y}{i(\omega \pm \omega^\star_{\rm c})}~,
\end{equation}
and
\begin{widetext}
\begin{equation}\label{eq:xzero}
 X_0=e \vD \left\{ (\gamma-1)g\frac{e}{2\hbar c} \frac{B}{p^2}+
\delta  \frac{\displaystyle \left[1-\gamma g\frac{e}{\hbar c}
 \frac{p_z}{p^3} B  + (2\gamma-1) \left( \frac{e}{2\hbar c} \right)^2 \frac{B^2}{p^4} \right] 
 }{\displaystyle 1-g\frac{e}{2\hbar c} \frac{p_z}{p^3} B } \frac{p_z}{p} 
 \right\} \frac{\tilde{E}_z}{i\omega}~,
 \end{equation}
 \end{widetext}
with
\begin{equation}\label{eq:wc}
 \omega^\star_{\rm c} =\omega^\star_{\rm c}(p,p_z)\equiv   \omega_{\rm c}\frac{\displaystyle 1-\gamma g\frac{e}{\hbar c} \frac{p_z}{p^3} B  }{\displaystyle 1-g\frac{e}{2\hbar c} \frac{p_z}{p^3} B  }~.
\end{equation}
In writing Eqs.~(\ref{eq:xpluminus})-(\ref{eq:xzero}) we have introduced another 
dimensionless control parameter, $\delta$. This takes two values: $\delta =1$ is what one should use, while $\delta = 0$ is what one should use to {\it artificially} discard the anisotropy in the distribution function.

The distribution function determines the total current carried by electrons at each Weyl node:
\begin{equation}
{\bm J}_g = -e \int \frac{ d^3 \bp}{(2 \pi)^3}  D_g(\bp) \dot{\br} f_g~.
\end{equation}
The factor $D_g(\bp)$ ensures that the number of states in the volume element remains constant in time~\cite{dixiao_rmp_2010}. For the sake of convenience, we decompose the current density per node in the sum of three terms: ${\bm J}_g={\bm J}^{(0)}_{g,E}+{\bm J}^{(0)}_{g,B}+\delta {\bm J}_{g}$, where
${\bm J}^{(0)}_{g,\lambda} = -e (2\pi)^{-3}\int d^3 \bp D_g(\bp)  \dot{\br}_{\lambda} f_g^{(0)}$ with $\lambda=E,B$ and 
$\delta {\bm J}_{g} = -e (2\pi)^{-3} \int d^3 \bp D_g(\bp)  \dot{\br}_{B} \delta f_g$.

We first examine in detail the dependence of ${\bm J}^{(0)}_{g,\lambda}$ on the static magnetic field ${\bm B}$.
In the weak magnetic field limit, we expand the distribution function $f^{(0)}_g$ in powers of ${\bm B}$, up to second order:
\begin{eqnarray}
  f_g^{(0)}[\varepsilon_{g}(\bp)] &\approx&   f^{(0)}_{g}[\varepsilon_{0}(\bp)]-
  \frac{\partial f^{(0)}_{g}}{\partial \varepsilon_g}\bigg|_{\varepsilon_{0}(\bp)} \gamma {\bm m}_g(\bp) \cdot {\bm B} \nonumber\\
  &+& \frac{1}{2}\frac{\partial^2 f^{(0)}_{g}}{\partial \varepsilon^2_g} \bigg|_{\varepsilon_{0}(\bp)}  [\gamma {\bm m}_g(\bp) \cdot {\bm B}]^2~.
\end{eqnarray}
Taking the limit $T \ll \epsF/k_{\rm B}$ we find
\begin{equation}\label{eq:J0_E}
{\bm J}^{(0)}_{g,E}=\gamma \frac{e^3 \vD}{24 \pi^2\hbar c \epsF}   {\bm B} \times \tilde{\bm E}(\omega) e^{-i \omega t}
\end{equation}
and
\begin{equation}\label{eq:J0_B}
 {\bm J}^{(0)}_{g,B}=g \frac{e^2 \epsF}{4 \pi^2 \hbar^2 c} {\bm B}~.
\end{equation}
We see that ${\bm J}^{(0)}_{g,E}$ is a) independent of the chirality $g$ of the Weyl node, and therefore leads to a finite correction to the ordinary Hall conductivity---see Eq.~(\ref{eq:Hall}) below---and b) proportional to the dimensionless parameter $\gamma$. Because of b), Eq.~(\ref{eq:J0_E}) originates from the orbital magnetic moment ${\bm m}_g({\bm p})$. The term ${\bm J}^{(0)}_{g,B}$ in Eq.~(\ref{eq:J0_B}) is proportional to the chirality $g$ of the Weyl node, and therefore has no effect on the total current ${\bm J} = \sum_g {\bm J}_g$ but yields a finite {\it axial} current ${\bm J}_{\rm ax}=\sum_g g {\bm J}_g$, in agreement with Refs.~\onlinecite{metlitski_prd_2005,jianhui_cpl_2013}.

We then evaluate the quantity $\delta {\bm J}_{g}$ and obtain the optical conductivity tensor $\sigma_{m n}(\omega)$. We first consider $\delta J_{g, z}$. By retaining all terms of second order in the ratio $\hbar\omega_{\rm c}/\mu$, we find~\cite{tau}
\begin{equation}\label{eq:szz_gd}
\sigma_{zz}(\omega) = i \frac{{\cal D}}{\pi \omega} {\cal C}_{zz}~, 
\end{equation}
where ${\cal D}=\pi e^2 n_{\rm e}/ m_{\rm c}$ is the Drude weight, 
$n_{\rm e}=\epsF^3/(3 \pi^2 \hbar^3  \vD^3)$ is the electron density, and 
\begin{equation}\label{eq:correction}
 {\cal C}_{zz}= \delta+ \frac{3 \hbar^2 \omega_{\rm c}^2}{4\epsF^2} 
\bigg\{ 1-\frac{3 \delta}{5}+\frac{\gamma[2\gamma(\delta+5)+11\delta-25]}{15}\bigg\}~.
\end{equation}
Eqs.~(\ref{eq:szz_gd})-(\ref{eq:correction}) are the most important results of this Section. 
Setting $\delta=\gamma=1$, we obtain the desired result for the longitudinal conductivity in the presence of a weak magnetic field:
\begin{equation}\label{eq:szz}
\sigma_{zz}(\omega)=  i\frac{\cal D}{\pi \omega}\Bigg(1+\frac{1}{5}  \frac{\hbar^2 \omega_{\rm c}^2}{\epsF^2}  \Bigg)~.
\end{equation}
Because of the non-trivial dependence of $\omega^\star_{\rm c}$ on $\bp$ in Eq.~(\ref{eq:wc}), the calculation
of $\delta J_{g, x}$, $\delta J_{g, y}$ at an arbitrary frequency $\omega$ is not straightforward. 
This calculation, however, notably simplifies in the low-frequency $\omega \ll \omega_{\rm c}$ limit, which is relevant for helicons. In this limit and after setting $\gamma  = \delta = 1$, we find
\begin{equation}
\sigma_{ xx}= \sigma_{yy}\approx
 - i \frac{ {\cal D} \omega  }{\pi \omega_{\rm c}^2 } \Bigg(1-\frac{1}{20}  \frac{\hbar^2 \omega_{\rm c}^2}{\epsF^2} \Bigg)~,
\end{equation}
and
\begin{equation}\label{eq:Hall}
\sigma_{ xy}= -\sigma_{yx}\approx
 \frac{{\cal D} }{\pi \omega_{\rm c}  }\Bigg(1+\frac{3}{20}   \frac{\hbar^2 \omega_{\rm c}^2}{\epsF^2} \Bigg)~.
\end{equation}
The remaining off-diagonal elements of the optical conductivity tensor, such as $\sigma_{xz}$, $\sigma_{yz}$, etc., vanish identically for symmetry reasons, independently of the frequency $\omega$. 

\noindent {\it Helicons and plasmons in WSMs.---}Using the wave equation (\ref{eq:wave}) and the semiclassical result for the optical conductivity tensor $\sigma_{mn}(\omega)$, 
we seek for collective modes of doped WSMs subject to a weak static magnetic field.  
To this aim, it is useful to introduce the dielectric tensor
\begin{equation}\label{eq:dielectric_tensor}
 \varepsilon_{\ell m} =\delta_{\ell m}\epsilon_{\rm b}+\frac{4 \pi i}{\omega} 
 \bigg[   \sigma_{\ell m} - \epsilon_{\ell m n}  \frac{ \alpha c}{2\pi^2} \Big(b_n - q_n \frac{b_0}{\omega} \Big)
 \bigg]
 \end{equation}
and ${\cal M}_{\ell m}= c^2 (q^2 \delta_{\ell m}-q_{\ell} q_m) - \omega^2 \varepsilon_{\ell m}$. In Eq.~(\ref{eq:dielectric_tensor}), $\epsilon_{\ell m n} $ is the 3D completely antisymmetric tensor and 
the Latin indices  $\ell$, $m$, and $n$ run over the Cartesian coordinates $x$, $y$, and $z$. Finally, a sum over $n$ is intended.

The zeroes of the determinant of ${\cal M}$ correspond to the self-sustained modes of a doped WSM. Following standard practice~\cite{Platzman_and_Wolff}, we focus on two special cases: i) $\bq$ parallel to the static magnetic field ${\bm B}$, i.e.~$\bq = q \hat{\bm z}$ and ii)  $\bq$ orthogonal to ${\bm B}$. When $\bq = q \hat{\bm z}$, four collective modes appear: three gapped modes, which are characterized by an energy of the order of Fermi energy, and a gapless mode, the helicon. If the wave vector $\bq$ is orthogonal to ${\bm B}$, 
we find only the three gapped modes, while the helicon solution is absent.  

After straightforward algebraic manipulations, we find the helicon dispersion relation in the long-wavelength limit: 
\begin{equation}\label{eq:helicon}
 \Omega_{\rm h} (q \to 0) = \frac{2 \alpha b_0 c q/\pi  + c^2 q^2}{\omega_{\rm p}^2/\omega_{\rm c} 
 + 2 \alpha c b_z/\pi}~,  
\end{equation}
where $\omega^2_{\rm p} = 4 \pi n_{\rm e } e^2 / m_{\rm c}$ is the 3D plasma frequency in a WSM. Eq.~(\ref{eq:helicon}) is the most important result of this Article. Note that Eq.~(\ref{eq:helicon}) is {\it independent} of the background dielectric constant $\epsilon_{\rm b}$~\cite{Platzman_and_Wolff}. 
Due to the time-like component $b_0$ of the axion angle, the helicon dispersion relation in a WSM is {\it linear} in $q$ rather than quadratic, the latter functional dependence on $q$ being the one occurring in  
ordinary metals---see Eq.~(\ref{eq:standard_helicon}).
Even for $b_0 = 0$, the helicon frequency differs from the textbook result (\ref{eq:standard_helicon}), in that the effective helicon mass $m_{\rm h}$ in a WSM depends
on the component of ${\bm b}$ along the direction of the static magnetic field ${\bm B}$: $m_{\rm h}= \hbar \omega_{\rm p}^2/(2 \omega_{\rm c}c^2) + \alpha \hbar b_z/(\pi c)$.

Before concluding, we comment on the gapped collective modes. For the sake of simplicity, we set ${\bm B} = {\bm 0}$ in the following analysis. In the long-wavelength limit, we find that the three gapped modes
$\Omega_{{\rm p}, \lambda}(q)$ with $\lambda = 1,2,3$ are given by:
\begin{equation}\label{eq:plasmons_partial}
\left\{
\begin{array}{l}
{\displaystyle \Omega_{{\rm p}, 1}(q=0)=\omega_-}\\ 
{\displaystyle \Omega_{{\rm p}, 2}(q=0)=\omega_{\rm p}/\sqrt{\epsilon_{\rm b}}}\\ 
{\displaystyle \Omega_{{\rm p}, 3}(q=0)=\omega_+}
\end{array}
\right.~,
\end{equation}
where $\omega_\pm = \sqrt{(\alpha c b)^2/(\pi\epsilon_{\rm b})^2  +  \omega^2_{\rm p}/\epsilon_{\rm b}}
\pm \alpha c b/(\pi\epsilon_{\rm b})$, with $b = |{\bm b}|$.
Very interestingly, we find that, unlike in an ordinary non-topological metal~\cite{Platzman_and_Wolff}, the degeneracy of the three gapped collective modes at $q = 0$ is {\it lifted} by the presence of the axion term ${\cal L}_{\theta}$ in the electromagnetic response. This is due to the fact that WSMs are optically gyrotropic media~\cite{zyuzinv arxiv 2014} with gyrotropy parameter proportional to $b = |{\bm b}|$.
Since the energy of the gapped collective modes is comparable to the Fermi energy, an accurate description of these modes requires the inclusion of the inter-band contribution $\sigma^{\rm inter}_{\ell \ell} (\omega)$~\cite{lv_ijmpb_2013,zhou_prb_2015,hoffmann_prb_2015} to the optical response, which has been neglected so far in our semiclassical approach:
\begin{equation}\label{eq:sigma_inter_band}
\sigma^{\rm inter}_{\ell \ell} (\omega) =\frac{\alpha c \omega}{12 \pi \vD} \left[\Theta(\hbar \omega - 2 \epsF) 
- \frac{i}{\pi} 
\log \left| \frac{4 \Lambda^2}{4 \epsF^2-\hbar^2 \omega^2} \right| \right]~.
\end{equation}
Here, $\ell = x,y,z$ and $\Lambda$ is an ultraviolet cut-off.
\begin{figure}[t]
\centering
\begin{overpic}[width=\columnwidth]{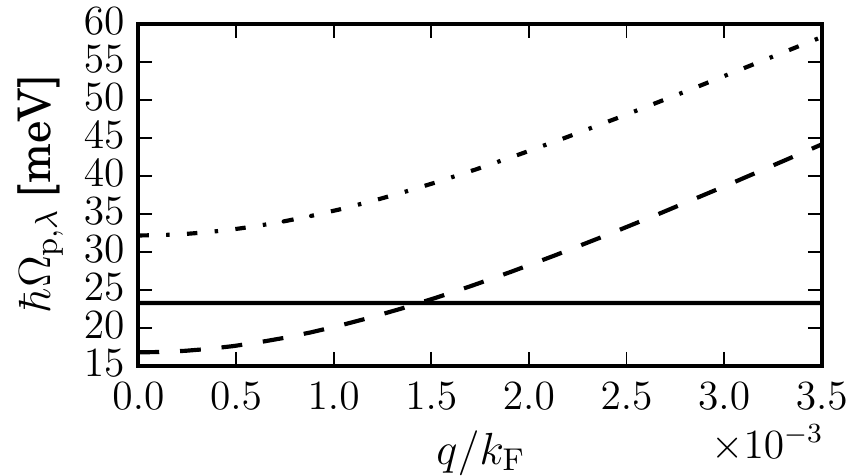}\put(2,52){\normalsize (a)}\end{overpic}\vspace{0.5em}
\begin{overpic}[width=\columnwidth]{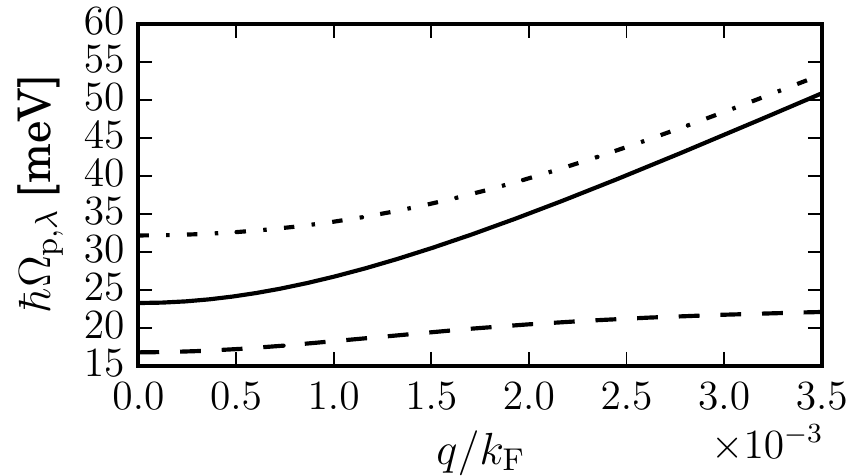}\put(2,52){\normalsize (b)}\end{overpic}
\caption{Dispersion relations $\hbar \Omega_{{\rm p}, \lambda}(q)$ (in meV) of collective gapped (``plasmon'') modes in doped Weyl semimetals as functions of wave vector $q$, in units of $k_{\rm F} = \epsF/(\hbar v_{\rm D})$. Such dispersions have been obtained by i) generalizing Eq.~(\ref{eq:plasmons_partial}) to finite $q \ll k_{\rm F}$ and ii) including inter-band effects, as from Eq.~(\ref{eq:sigma_inter_band}). 
Dashed line: $\lambda=1$. Solid line: $\lambda=2$. Dash-dotted line: $\lambda=3$.   
Panel (a): results for ${\bm q}$ parallel to ${\bm b}$. Panel (b): results for ${\bm q}$ orthogonal to ${\bm b}$. 
Results in this figure have been obtained by setting $\varepsilon_{\rm b}=5$,   $\epsF=40~{\rm meV}$,  
$\vD=c/1000$,  $b_0=0$, ${\bm b}=(0.01,0.01,0.01)\pi/a$, and $\Lambda= \hbar \vD \pi/a$ with $a=3.5$~\AA.
This choice of microscopic parameters is justified by recent experimental results in NbAs and TaAs~\cite{lv_prx_2015,lv_arxiv_2015a, xu_prx_2015}.
\label{fig:modes}} 
\end{figure}
Fig.~\ref{fig:modes} shows the dispersion relations $\Omega_{{\rm p}, \lambda}(q)$ of the $\lambda=1,2,3$ gapped collective modes, as calculated by adding $\sigma^{\rm inter}_{\ell \ell} (\omega)$ to the intra-band semiclassical contribution $\sigma_{xx}(\omega) = \sigma_{yy}(\omega) =\sigma_{zz}(\omega)$, i.e.~Eq.~(\ref{eq:szz}) evaluated at ${\bm B} = {\bm 0}$ and for $\gamma = \delta = 1$. In Fig.~\ref{fig:modes}(a) [(b)] the wave vector $\bq$ is parallel [orthogonal] 
to the space-like component ${\bm b}$ of the axion angle. The main effect of the inter-band contribution on the collective modes is to redshift their gaps at $q=0$. This can be easily explained by recognizing that, in the long-wavelength limit, the inter-band contribution $\sigma^{\rm inter}_{\ell \ell} (\omega)$ to the optical response can be described, to a very good approximation, as a renormalization of the background dielectric constant, i.e.~$\epsilon_{\rm b} \to \epsilon_{\rm b} + \alpha c \log[|4 \Lambda^2/(4 \epsF^2-\hbar^2 \omega_{\rm p}^2/\epsilon_{\rm b})|]/(3 \pi \vD)$.
In Fig.~\ref{fig:modes}(a) we find two {\it transverse} modes with a quadratic dispersion relation and a dispersionless longitudinal mode. We note that in this case there is no mixing of the two transverse modes with the longitudinal mode, exactly like in an ordinary metal. What is peculariar to WSMs is that when $\bq$ is tilted away from ${\bm b}$, 
the lowest-energy transverse mode hybridizes with the longitudinal mode. This effect is maximal for $\bq$ orthogonal to ${\bm b}$, as in Fig.~\ref{fig:modes}(b). 

In summary, we have evaluated the optical conductivity tensor of a 3D Weyl semimetal from semiclassical Boltzmann transport theory, with the inclusion of the orbital moment ${\bm m}_g(\bp)$ and anisotropic contributions to the distribution function. A general expression for the longitudinal conductivity is reported in Eq.~(\ref{eq:szz_gd}). We have used the calculated optical conductivity tensor together with the axion contribution (\ref{eq:axionterm}) to the standard electromagnetic Lagrangian to find the collective modes of a 3D Weyl semimetal. We have demonstrated that the axion term dramatically alters the helicon dispersion, Eq.~(\ref{eq:helicon}), with respect to that in non-topological metals, Eq.~(\ref{eq:standard_helicon}). Finally, we have highlighted axion anomalies in the gapped sector of collective excitations, Fig.~\ref{fig:modes}, by taking into account inter-band corrections to the semiclassical (intra-band) optical response.

\noindent {\it Acknowledgements.---}This work was supported by a 2012 Scuola Normale Superiore internal project (F.M.D.P.), the EC under the Graphene Flagship program (contract no.~CNECT-ICT-604391) (M.I.K. and M.P.), MIUR through the program ``Progetti Premiali 2012'' - Project ``ABNANOTECH'' (M.P.), 
the European Research Council (ERC) through the Advanced Grant No. 338957 FEMTO/NANO (M.I.K.), 
and NWO via the Spinoza Prize (M.I.K.). Free software (www.gnu.org, www.python.org) was used.

\end{document}